\documentclass[aps,prl,twocolumn,amsmath,amssymb,superscriptaddress,10pt]{revtex4-2}

\usepackage{comment}
\usepackage{xcolor}
\usepackage{graphicx}
\usepackage{dcolumn}
\usepackage{bm}
\usepackage{hyperref}
\usepackage{mathtools}
\usepackage[mathlines]{lineno} 

\usepackage{cleveref}

\usepackage{tikz}
\usetikzlibrary{shapes.geometric}
\newcommand{\hexagon}[1][.1pt]{%
  \begin{tikzpicture}[baseline=-0.5ex]
    \pgfmathsetmacro{\s}{#1}
    \draw[line width=0.8pt]
      (\s,0)
      \foreach \a in {60,120,180,240,300,360} { -- ({\s*cos(\a)},{\s*sin(\a)}) }
      -- cycle;
  \end{tikzpicture}%
}

\graphicspath{{PaperFiguresFinal}}


\begin{document}

\title{Hearing the light: stray-field noise from the emergent photon in quantum spin ice}

\author{Gautam K. Naik}
\thanks{These authors contributed equally to this work.}
\affiliation{Department of Physics, Boston University, Boston, Massachusetts 02215, USA}

\author{Jonathan N. Hall\'en}
\thanks{These authors contributed equally to this work.}
\affiliation{Department of Physics, Boston University, Boston, Massachusetts 02215, USA}
\affiliation{Department of Physics, Harvard University, Cambridge, Massachusetts 02138, USA}

\author{Nishan C. Jayarama}
\affiliation{Department of Physics, Boston University, Boston, Massachusetts 02215, USA}

\author{Roderich Moessner}
\affiliation{Max-Planck-Institut f\"{u}r Physik komplexer Systeme, 01187 Dresden, Germany}

\author{Chris R. Laumann}
\affiliation{Department of Physics, Boston University, Boston, Massachusetts 02215, USA}
\affiliation{Department of Physics, Harvard University, Cambridge, Massachusetts 02138, USA}
\affiliation{Max-Planck-Institut f\"{u}r Physik komplexer Systeme, 01187 Dresden, Germany}

\begin{abstract}

Decisive experimental confirmation of the $U(1)$ quantum spin liquid phase in quantum spin ice remains an outstanding challenge. 
In this work, we propose stray-field magnetometry as a direct probe of the emergent photons -- the gapless excitation of the emergent electrodynamics in quantum spin ice.  
The emergent photons are transverse magnetization waves, which, in a finite sample, form discrete modes governed by one of two sets of natural boundary conditions: ``insulating'' or ``superconducting''. 
Considering cavity and thin film geometries, we find that the spectrum and spatial structure of the stray magnetic noise provide a sharp qualitative signature of the underlying electrodynamics.  
The predicted stray-field noise power lies comfortably within the detection range of present-day solid-state defect magnetometry.  

\end{abstract}

\maketitle


The family of frustrated quantum magnets known as \emph{quantum spin ice} (QSI) is predicted to realize a $U(1)$ gauge theory analogous to quantum electromagnetism and commonly referred to as the Coulomb phase \cite{moessnerThreedimensionalResonatingvalencebondLiquids2003,huseCoulombLiquidDimer2003,hermelePyrochlorePhotons$U1$2004,banerjeeUnusualLiquidState2008,shannonQuantumIceQuantum2012,bentonSeeingLightExperimental2012,savaryQuantumSpinLiquids2017,knolleFieldGuideSpin2019,udagawaSpinIce2021}.
A defining property of the Coulomb phase is the presence of long-wavelength, gapless excitations in the form of transverse magnetization waves -- the \emph{emergent photon}. 
A range of rare-earth pyrochlore materials have been proposed as QSI candidates \cite{gingrasQuantumSpinIce2014,rauFrustratedQuantumRareearth2019}, and advanced synthesis techniques allow for the experimental study of both high quality bulk crystals \cite{udagawaSpinIce2021} and thin films \cite{bovoRestorationThirdLaw2014, leusinkThinFilmsSpin2014, xingProbingIceRuleBreakingTransition2025}.
The analogy between the emergent photon and real light suggests that controlling the sample geometry can be turned to experimental advantage through ``cavity'' and/or ``waveguide'' engineering.

%
%
\begin{figure}[b]
    \centering
    \includegraphics[width=\linewidth]{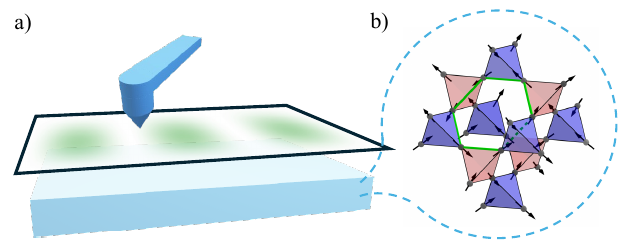}
    \caption{Stray-field magnetic noise from quantum spin ice. a) We consider a finite sample of QSI, and probe the stray-field magnetic noise with a scanning magnetometer above it. b) Structure of QSI on the pyrochlore lattice. The strong ZZ interactions force the spins to be in 2-in-2-out ``ice'' configurations \cite{udagawaSpinIce2021}. Flipping spins aligned along hexagonal rings (green) connect the ice configurations.}
    \label{fig:Fig1}
\end{figure}
%
%

Much of the current effort to observe QSI physics centers on the experimental detection of the emergent photons. 
While recent experimental results on the Cerium pyrochlores are consistent with Coulomb phase physics \cite{gaudetQuantumSpinIce2019,gaoExperimentalSignaturesThreedimensional2019,bhardwajSleuthingOutExotic2022,smithCase$mathrmU1_ensuremathpi$Quantum2022,zhaoInelasticNeutronScattering2024,gaoNeutronScatteringThermodynamic2025,poreeEvidenceFractionalMatter2025}, conclusive direct detection of the photon remains an open challenge; the photon is expected to appear at low frequencies ($\lesssim 1$~GHz), only exists at low temperatures ($\lesssim 0.5$~K), and has no weight at zero momentum, making it very difficult to detect with common probes like neutron scattering or bulk susceptibility measurements. 

Stray-field magnetic noise spectroscopy offers a promising alternative route to observe the emergent photons, since they carry transverse magnetization which generates spatiotemporally structured stray magnetic fields. 
There are two primary approaches for such measurements. 
The development of magnetic noise sensing using solid-state color centers now allows for local, highly sensitive magnetic noise measurements at frequencies from DC to the GHz regime \cite{hongNanoscaleMagnetometryNV2013,luanDecoherenceImagingSpin2015,degenQuantumSensing2017,casolaProbingCondensedMatter2018,chatterjeeDiagnosingPhasesMagnetic2019,rovnyNanoscaleDiamondQuantum2024,perskyStudyingQuantumMaterials2022,khooProbingQuantumNoise2022,kirschnerProposalDetectionMagnetic2018}.
Alternatively, scanning tip SQUID magnetometers typically operate at somewhat lower frequencies, although there have been demonstrations of similar noise sensitivities up to 100 MHz with SQUIDs \cite{perskyStudyingQuantumMaterials2022,cuiScanningSQUIDSampler2017,foroughiMicroSQUIDDispersiveReadout2018,wyssMagneticThermalTopographic2022,kirtleyScanningSQUIDSusceptometers2016}, and larger SQUID magnetometers have been used to measure global magnetization fluctuations in quantum magnets \cite{dusadMagneticMonopoleNoise2019,samarakoonAnomalousMagneticNoise2022,takahashiSpiralSpinLiquid2025}.
In either case, the physical quantity probed is the tensorial magnetic noise spectral density at the position $\vec{r}$ of the probe,
\begin{equation}
    {\cal C}^{\mu \nu}(\vec{r}, \omega) = \int_{-\infty}^\infty {\rm d} t\, e^{i\omega t} \left\langle \left\{B^\mu (\vec{r}, 0), B^\nu(\vec{r}, t) \right\} \right \rangle 
    \, ,
    \label{eq:MagNoise}
\end{equation}
%
where $\vec{B}(\vec{r}, t)$ is the (true) magnetic field, $\langle\ \rangle$ denotes a thermal ensemble average and $\{\cdot ,\cdot \}$ denotes the anti-commutator \cite{chatterjeeDiagnosingPhasesMagnetic2019}. 
(We use capitalized $\vec{E}$ and $\vec{B}$ to refer to the true electromagnetic fields and lower-case $\vec{e}$ and $\vec{b}$ for the emergent.)

The low-frequency dynamics of the Coulomb phase are described by the Maxwell action,
\begin{equation}
    {\cal S} = \int {\rm d}t \int_V {\rm d}^3 x\, \frac{\hbar}{8\pi \alpha' v} \left(e^2 - v^2 b^2 \right) 
    \label{eq:Action_Maxwell}
\end{equation}
where $V$ is the sample volume. 
The $\vec{e}$ field carries the physical magnetization of the emergent photon, whose fluctuations induce the magnetic field noise in Eq.~\eqref{eq:MagNoise}. 
The Coulomb phase is characterized by two couplings -- the dimensionless fine-structure constant $\alpha'$ (typically $\approx 1/10$~\cite{paceEmergentFineStructure2021}) and the speed of light $v$.
The latter has been estimated as $v\approx10$~m/s~\cite{hermelePyrochlorePhotons$U1$2004,sibilleExperimentalSignaturesEmergent2018,bentonSeeingLightExperimental2012,shannonQuantumIceQuantum2012,kwasigrochSemiclassicalApproachQuantum2017}; if one probes the magnetic noise around 1~MHz, spatial structure thus appears at length-scales of about 10~$\mu$m -- the wavelength of the 1~MHz photons.
Furthermore, this spatial structure can be resolved by a noise probe at micrometer distances from the sample (see Fig.~\ref{fig:Fig1}), well within experimental feasibility.

We show that finite-size quantization leads to sharp spectral features as well as extended spatial noise patterns, which are strongly characteristic of the existence of an emergent photon.
Whether this noise is observable through stray-field magnetometry is, however, dependent on the effective long-wavelength boundary conditions on the Maxwell theory.

\paragraph{Boundary conditions of QSI.}
To understand the effect of finite-size geometry, we first need to establish the appropriate long-wavelength boundary conditions for the emergent electromagnetic fields. 
The most obvious point is that these fields do not exist outside the material. 
Furthermore, we assume that any other degrees of freedom at the boundary are gapped; gapless phases of matter in two dimensions are rare and typically appear only for fine-tuned theories (in the absence of fermions).
Furthermore, in QSI, we expect the energy scales at the boundary to be large compared to those governing the bulk physics, as will be explained further below.
There are thus no degrees of freedom for long-wavelength emergent photons to couple to at the boundary, and we expect them to reflect elastically.

To derive natural boundary conditions, we make the following formal assumptions:
(i) there is no energy transfer through or into the boundary;
(ii) the entire system is time-reversal symmetric;
and (iii) the wavelength of the photon modes is much larger than any microscopic length scale at the boundary.
Under these assumptions, we obtain two types of homogeneous boundary conditions:
\begin{align}
    \vec{b}_\parallel&=\vec{0} \ {\rm and} \ e_\perp=0 \quad \text{ (``insulating'')}
    \label{eq:insulating_bc}\\
    \vec{e}_\parallel&=\vec{0}  \ {\rm and} \  b_\perp=0   \quad \text{ (``superconducting'')} 
    \label{eq:superconducting_bc}
\end{align}
We refer to these as ``insulating'' and ``superconducting'', by analogy to the corresponding conditions in true electromagnetism. 

We can understand the boundary conditions as follows:
By (i), the normal component of the Poynting vector has to vanish at the boundary, $(\vec{e}\times \vec{b})_{\perp}=0$.
This requires the parallel components of the fields to satisfy either $\vec{e}_{\parallel}\propto \vec{b}_{\parallel}$ or $\left\vert\vec{e}_{\parallel}\right\vert \left\vert \vec{b}_{\parallel}\right\vert=0$.
The former is not allowed: (ii) disallows boundary conditions of the form $\vec{e}_{\parallel}=u \vec{b}_{\parallel}$, where $u$ is a velocity, and (iii) disallows boundary conditions involving spatial or temporal derivatives.
This leaves two possibilities.
If $\vec{b}_{\parallel}=\vec{0}$, there is no boundary current and, by continuity at finite frequency, no charge; this gives $e_{\perp}=0$ and hence the ``insulating'' boundary condition.
$\vec{e}_{\parallel}=\vec{0}$ along with the homogeneous Maxwell equation, implies $b_{\perp}=0$ and hence the ``superconducting'' boundary condition. 

One can see how these boundary conditions appear in practice by considering a minimal phenomenological model.
Take a system where the emergent fields couple to gapped bosonic matter at the boundary, as described by a complex scalar field with the action
\begin{align}
    S_{\rm bdy} = \int {\rm d}t& \int_{\partial V} {\rm d}^2x \,\hbar\, \bigg\{ |D_t \psi|^2  \nonumber\\
    & -u^2|D_i \psi|^2- m^2 |\psi|^2 - \frac{\Lambda}{2} |\psi|^4\bigg\}
    \, ,
    \label{eq:Action_boundary}
\end{align}
where $D_{\mu}=\partial_{\mu}+ia_{\mu}$.
Computing the equations of motion from the total action, Eq.~\eqref{eq:Action_Maxwell} and Eq.~\eqref{eq:Action_boundary}, one immediately arrives at the insulating boundary conditions when the mass term is $m^2>0$. 
When $m^2<0$, we arrive at the superconducting boundary conditions in the limit that the electric and magnetic screening lengths at the boundary are much shorter than the wavelength of the photon modes
 \cite{SM}. 

The ideal, long-wavelength boundary conditions, Eqs.~\eqref{eq:insulating_bc} and \eqref{eq:superconducting_bc}, provide a natural reference point for our analysis. 
If the assumptions (i-iii) break down, cavity quantization persists, but with modified spectra and finite broadening due to energy loss into the boundary.
General microscopic considerations suggest that well-gapped boundary conditions are natural for QSI, as will be discussed further at the end of this paper.

\paragraph{QSI as an electromagnetic cavity.}
A finite sample of QSI should act as an electromagnetic cavity independent of which boundary condition applies.
We can expand the vector potential $\vec{a}$ in a discrete set of cavity modes (in the temporal gauge, $\phi = 0$)
\begin{equation}
    \hat{\vec{a}}\left(\vec{r}, t \right) = 
     \sum_s \sqrt{\frac{4\pi \alpha' v}{V \omega_s}} \vec{W}_s(\vec{r}) 
    \left[\hat{d}_s e^{-i\omega_s t} + \hat{d}^\dagger_s e^{i\omega_s t}\right] 
    \, .
    \label{eq:modes}
\end{equation}
$\hat{d}_s$ and $\hat{d}^\dagger_s$ are bosonic raising and lowering operators for the photon modes with frequency $\omega_s$. 
$\vec{W}_{s}(\vec{r})$ is a geometry and boundary condition-dependent, order-1, dimensionless vector field which encodes the polarization and amplitude of the mode in space \cite{SM}. 
In cuboid geometries, the modes are naturally parametrized by the vector $\vec{k}= \pi (n_x/L_x, n_y/L_y, n_z/L_z)$, with $n_{x, y, z}\in {\mathbb Z}^+$ and frequency $\omega = v \left\vert \vec{k}\right\vert$.
As in conventional optical cavities, not all combinations of $n_x$, $n_y$, $n_z$ are allowed: At least two components must be non-zero, and if all three components are non-zero, there are two independent polarizations available \cite{SM}.
Unlike conventional cavities, the particular allowed polarizations depend on the choice of insulating or superconducting boundary conditions.

%
%
\begin{figure}[ht]
    \centering
    \includegraphics[width=\linewidth]{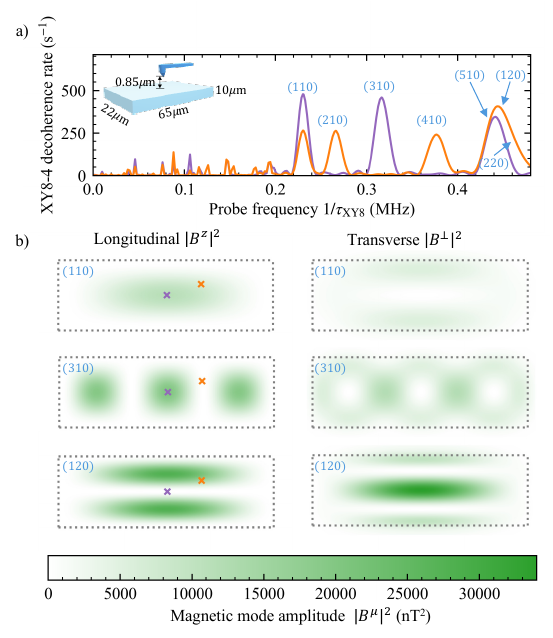}
     \caption{Magnetic noise generated by a finite quantum spin ice sample with ``superconducting'' boundary conditions at temperature 100~mK. 
     a) $T_2$ decoherence time of an NV center in the magnetic stray-field outside the sample (geometry in inset) for an XY8-$4$ dynamical decoupling protocol, plotted versus the pulse spacing $\tau_{\rm XY8}$. The $T_2$ time is computed by convolving the longitudinal component of the discrete magnetic noise spectrum with the XY8-$N$ filter function \cite{degenQuantumSensing2017,rovnyNanoscaleDiamondQuantum2024, SM}. The curves show the $T_2$ times at two different points above the sample, as indicated in panel (b).
     b) Spatially resolved magnetic noise magnitude for a selection of low-frequency modes labeled $(n_x, n_y, n_z)$, with the longitudinal component ($|B^z|^2$) on the left and the transverse component ($|B^\perp|^2=|B^x|^2+|B^y|^2$) on the right, and mapped out in a plane 0.85~$\mu$m above the sample surface (see inset in (a)).
     Note that for the ``insulating'' boundary conditions, the stray-field noise is exactly zero everywhere outside the sample.
     Converted to experimentally relevant units using $v=10$~m/s, $\alpha'=0.1$, and $\mu_0^2 g^2 = 10^{-38}\ {\rm T^2 m^{-2} s^2}$.}
    \label{fig:Fig2}
\end{figure}
%
%

At finite temperature, thermal occupation of the normal modes produces well quantized magnetic noise spectra (see Fig.~\ref{fig:Fig2}a).
%
The magnetic noise tensor [Eq.~\eqref{eq:MagNoise}] can be expanded in terms of the modes as ${\cal C}^{\mu \nu}(\vec{r}, \omega) = \sum_s B_s^\mu(\vec{r})B_s^\nu(\vec{r}) [2n_{\rm B}(\omega_s) + 1] 2\pi \delta(\omega-\omega_s)$, where $n_{\rm B}(\omega)=[\exp{(\beta \hbar \omega)}-1]^{-1}$ is the Bose-Einstein occupation function.
$B_s^\mu(\vec{r})$ is the zero-point magnetic field generated at position $\vec{r}$ by mode $s$. 
It is computed by convolving the magnetization $\vec{M}=g \vec{e}$ with the dipole kernel $H^{\mu\nu}(\vec{R})$ over the sample volume $V$:
\begin{align}
    \nonumber
    B_s^\mu(\vec{r}) &=\mu_0 g \sqrt{\frac{4\pi \alpha' v  \omega_s}{V}} \sum_\nu \int_V {\rm d}^3 r' H^{\mu \nu}(\vec{r}-\vec{r}') W^\nu_s (\vec{r}') 
    \, ,\\
    &H^{\mu \nu}\left(\vec{R}\right)
    =\frac{1}{4\pi}\
     \frac{\delta^{\mu \nu}R^{2}-3R^\mu R^\nu}{R^{5}}
     \, .
     \label{eq:B_dipole_kernel_integral}
\end{align}

Strikingly, the photon modes in a sample with insulating boundary conditions generate no stray-field magnetic noise at all.
This can be intuitively understood in the Gilbert model: the insulating boundary condition does not permit ``magnetic charge'' ($\vec{M}\cdot\hat{n}=0$) at the surface, which along with the bulk Gauss law ($\vec{\nabla}\cdot\vec{M}=0$), means there are no sources for stray magnetic fields outside the sample.
This can also be formally shown by performing the kernel integral [Eq.~\eqref{eq:B_dipole_kernel_integral}] in reciprocal space \cite{SM}.
In contrast, the superconducting boundary conditions permit surface magnetic charge and hence, stray magnetic fields.
In this case, we perform the integral in Eq.~\eqref{eq:B_dipole_kernel_integral} for generic, fully finite geometries using Ewald summation to handle the slowly converging integral \cite{ewaldBerechnungOptischerUnd1921,SM,code}.

Focusing on the superconducting boundary conditions, the cavity quantization of the emergent photon modes manifests as sharp features in the stray-field magnetic noise.
In Fig~\ref{fig:Fig2}a we illustrate this through the magnetic noise induced XY8-4 decoherence time of a diamond nitrogen-vacancy center (NV center) placed above the QSI sample \cite{degenQuantumSensing2017,rovnyNanoscaleDiamondQuantum2024, SM}.
Besides sharp features in frequency, there is also mode-dependent spatial structure, which could be mapped using either a scanning probe or an array of probes. 
Fig~\ref{fig:Fig2}b provides examples of the spatial structure in the stray noise generated by specific modes, with further examples in the SM \cite{SM}.

%
\begin{figure}[b]
    \centering
    \includegraphics[width=\linewidth]{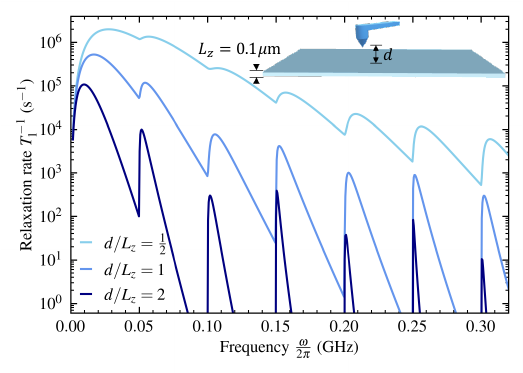}
    \caption{$T_1$ decoherence time of an NV center placed at different distances, $d$, from a thin film QSI with ``superconducting'' boundaries. 
    The decoherence time is directly determined from the stray-field magnetic noise: $T_1^{-1}=\frac{3}{2}\gamma_e^2(\mathcal{C}^{xx}+\mathcal{C}^{yy})$.
    The steps at frequencies $\omega = n \pi v/L_z$, appear as each additional longitudinal mode $n$ becomes available. 
    Converted to experimentally relevant units using $\gamma_e=2\pi \times 28$~GHz/T, $v=10$~m/s, $\alpha'=0.1$, and $\mu_0^2 g^2 = 10^{-38}\ {\rm T^2 m^{-2} s^2}$, and assuming a thermal photon population at temperature 100~mK.}
    \label{fig:Fig3}
\end{figure}
%
%

\paragraph{Thin film QSI as a waveguide.}
Another experimentally relevant sample class is pyrochlore thin films.
For thin films of thickness $L_z$ and infinite extent in $x$ and $y$, one can perform the dipole kernel integral analytically \cite{SM}.
The stray-field noise is again exactly zero for insulating boundary conditions.
The photon modes correspond to traveling waves in the transverse directions and standing waves in $z$, conveniently labeled by a transverse momentum $\vec{k}=(k_x, k_y)$ and an integer index $n_z\ge0$, with frequency $\omega_n\left(\vec{k}\right) = v \sqrt{k^2 + (\pi n_z/L_z)^2}$.
Fig.~\ref{fig:Fig3} shows the stray-field magnetic noise spectrum of such a thin film with superconducting boundary conditions at varying sample-probe distances $d$ -- expressed in terms of the magnetic noise induced depolarization rate $T_1^{-1}$ of an NV center \cite{degenQuantumSensing2017, rovnyNanoscaleDiamondQuantum2024}.
With increasing frequency, the mode density rises sharply at $\omega = n \pi v / L_z $ for integer $n$ -- i.e., when additional standing wave modes become available. 
This manifests as sudden increases in the noise spectral density at the corresponding frequencies. 
Magnetization modes produce stray fields decaying approximately as $k^2 \exp{\left(-2\left\vert\vec{k}\right\vert d\right)}$ at distance $d$ from the film \cite{chatterjeeDiagnosingPhasesMagnetic2019,SM} -- this accounts for the overall exponentially decaying trend of Fig.~\ref{fig:Fig3}.
It also explains why the peaks in the spectrum are sharper for larger $d$; at larger probe distances, the difference in suppression between small- and large-transverse-momentum modes becomes more pronounced.
Counterintuitively, it can therefore be easier to resolve structure in the spectrum using a more distant probe.

\begin{figure}
    \centering
    \includegraphics[width=\linewidth]{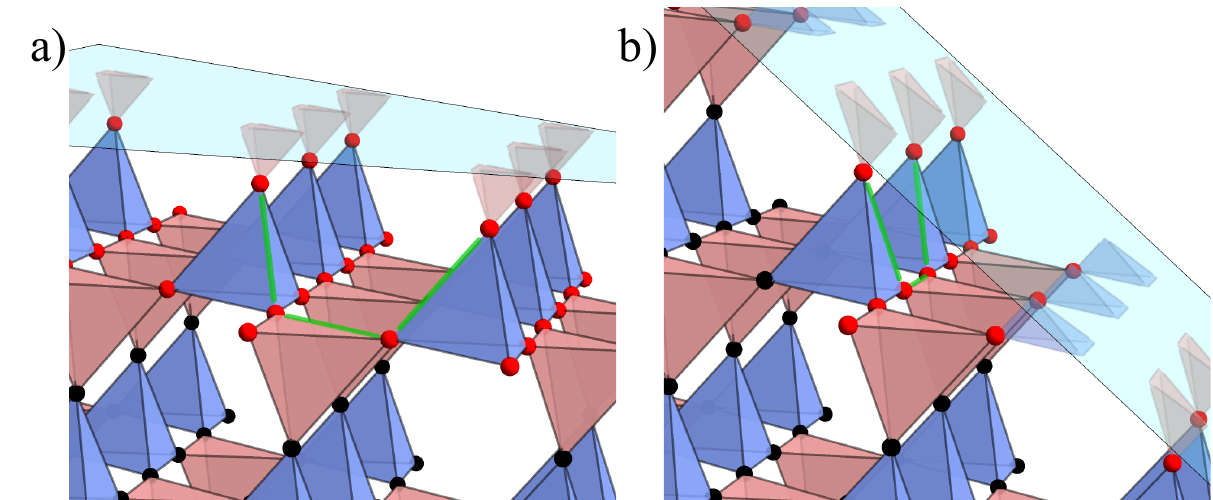}
    \caption{
    Termination of pyrochlore QSI with (a) $(111)$ plane and (b) $(110)$ plane. The broken tetrahedra host $e$-charges, which can hop along broken hexagons on the boundary (green paths). For these example terminations, the hopping terms are generated at second order in a Schrieffer-Wolff expansion and each involves four spins in the boundary layer (marked by red spheres). Note that first-order, two-spin terms would also appear at the boundaries if the termination planes were shifted to leave cut tetrahedra containing three spins.
    }
    \label{fig:Fig4}
\end{figure}

\paragraph{Microscopic QSI boundaries.}
The Coulomb phase is typically found in short-ranged Hamiltonians on the pyrochlore lattice with a dominant nearest-neighbor ZZ interaction
\begin{equation}
    {\cal H} = J_{ZZ} \sum_{\langle ij\rangle}  s^z_i s^z_j + {\cal H}'
    \, ,
    \label{eq:xxz_hamiltonian}
\end{equation}
where $J_{ZZ}>0$ and ${\cal H}'$ contains all other symmetry-allowed two-spin terms \cite{hermelePyrochlorePhotons$U1$2004,savaryQuantumSpinLiquids2017,rauFrustratedQuantumRareearth2019,naikThetaElectromagnetismQuantum2025}.
Without ${\cal H}'$, the ground-state manifold consists of an extensive set of degenerate 2-in-2-out ice states -- configurations that satisfy a lattice Gauss law with no bulk charges (see Fig.~\ref{fig:Fig1})
\begin{equation}
    q_t \equiv {\rm div}_t (s^z) = 
    \left\{
    \begin{aligned}
       0 \quad & {\rm bulk}\\
       \pm \frac{1}{2} \quad & {\rm boundary}
    \end{aligned}\right.
\end{equation}
where $t$ denotes tetrahedra in the lattice.
%
Terminating the pyrochlore lattice inevitably produces broken tetrahedra (see Fig.~\ref{fig:Fig4} for examples), and those containing an odd number of spins necessarily host degenerate boundary charge states in the ground state manifold \cite{lantagne-hurtubiseSpinIceThinFilms2018}.

In the usual derivation, the terms in ${\cal H}'$ lift the degeneracy of the bulk ice states and induce dynamics which ultimately lead to the emergent photons. 
At the boundary, such terms induce dynamics for the low-energy charges, which are nonetheless coupled to the bulk degrees of freedom through Gauss' law.
This is captured by our long-wavelength phenomenological model [Eq.~\eqref{eq:Action_boundary}].

In more detail, Schrieffer-Wolff projection of ${\cal H}'$ onto the ice manifold produces the well-known ring exchange Hamiltonian in the bulk \cite{hermelePyrochlorePhotons$U1$2004}.
\begin{equation}
    {\cal H}_{\rm bulk} = -J_{\rm ring} \sum_{{\hexagon}} (s_1^+s_2^-s_3^+s_4^-s_5^+s_6^- \quad + \quad {\rm h.c})
    \,.
    \label{eq:ringexchange}
\end{equation}
The sum runs over all minimal six cycles on the lattice, and the indices $1-6$ run over the spins on each cycle (see Fig.~\ref{fig:Fig1}b). 
The low-energy effective Hamiltonian contains additional terms at the boundary corresponding to open paths connecting free boundary charges (see Fig.~\ref{fig:Fig4}).
These terms hop the boundary charges, and whether the boundary realizes an insulating or superconducting phase depends on the details of these hopping terms, the geometry of the boundary, and the interplay with the bulk.

Since the boundary terms can correspond to paths with length less than six, they are generated at lower order in the Schrieffer-Wolff expansion than the bulk ring exchange term.
This leads to a separation of energy scales between the bulk and boundary, which we invoked when deriving the natural boundary conditions.

In general, there is no reason to expect crystal boundaries to correspond to simple symmetric planes like those shown in Fig.~\ref{fig:Fig4}; a given crystal can exhibit many distinct microscopic terminations. 
Attempting to enumerate and classify all such terminations is therefore a major undertaking, beyond the scope of this work.
Here, we have restricted our attention to the ideal long-wavelength boundary conditions that capture the universal, coarse-grained physics, independent of microscopic details.

%
%
%
%

\paragraph{Parameter choices and experimental feasibility.}
The boundary conditions and photon modes we predict apply independent of the specific QSI material realization, but in order to convert our results to experimentally relevant units, we have made some order-of-magnitude estimates of material-dependent parameters.
The parameters $v\approx 10$~m/s and $\alpha'\approx1/10$ have been estimated elsewhere \cite{sibilleExperimentalSignaturesEmergent2018,bentonSeeingLightExperimental2012,shannonQuantumIceQuantum2012,kwasigrochSemiclassicalApproachQuantum2017,paceEmergentFineStructure2021}. 
The coupling $g$ is fixed by the size of the microscopic moments, $g \approx \frac{2\mu}{v \alpha' l_0}$ \cite{castelnovoMagneticMonopolesSpin2008, laumannHybridDyonsInverted2023}. 
Taking as a quantitative example the Pr-based compounds (Pr$_2$Zr$_2$O$_7$ and Pr$_2$Hf$_2$O$_7$), we estimate a lattice scale $l_0\approx 0.3$~nm and spins with magnetic dipole moment $\mu \approx 2.5\mu_{\rm B}$ \cite{ciomagahatneanStructuralMagneticProperties2014}, and find $\mu_0g\approx 10^{-19}$~Tm$^{-1}$s.
Similar estimates apply to all three symmetry classes of QSI models \cite{rauFrustratedQuantumRareearth2019}, given that the dominant, ice-rule inducing interaction is between spin components which carry dipole moments. 
(The compounds Ce$_2$Zr$_2$O$_7$ and Ce$_2$Sn$_2$O$_7$ are believed to have a dominant interaction in the octupole sector and our estimates do not apply \cite{gaoExperimentalSignaturesThreedimensional2019,zhaoInelasticNeutronScattering2024,gaoNeutronScatteringThermodynamic2025,poreeEvidenceFractionalMatter2025}.)
We have computed noise spectra at 100~mK; for the main material candidates, this is in the temperature regime for which QSI physics are predicted \cite{gingrasQuantumSpinIce2014,rauFrustratedQuantumRareearth2019}, and it is also reachable with a typical dilution fridge.

For a finite geometry of approximate size $(100~{\rm \mu m})^3$, the lowest-energy emergent photon modes have frequencies of order 100~kHz -- a suitable regime for $T_2$ magnetometry with solid-state defects~\cite{degenQuantumSensing2017, rovnyNanoscaleDiamondQuantum2024}, as exemplified in Fig~\ref{fig:Fig2}.
The typical thickness of rare-earth pyrochlore thin films is about 100~nm~\cite{leusinkThinFilmsSpin2014,bovoRestorationThirdLaw2014,liuInsituFabricationTransport2020, liuMagneticWeylSemimetallic2021,wenEpitaxialStabilization111oriented2021, rabinovichEpitaxialGrowthStoichiometry2024, andersonDefectEngineeringEpitaxial2024}, and the stray-field noise in the 100~MHz to GHz regime should thus contain signatures of the photon mode structure.
This frequency regime is suitable for NV center $T_1$ magnetometry (``relaxometry'') ~\cite{degenQuantumSensing2017, rovnyNanoscaleDiamondQuantum2024}, as shown in Fig.~\ref{fig:Fig3}.
The thin film geometry also has the benefit that a few layers of a non-magnetic material can be grown on top to control the termination -- although it is worth noting that strain from the interface may lift the degeneracy of the ice states and induce ordering~\cite{bovoRestorationThirdLaw2014, xingProbingIceRuleBreakingTransition2025, bovoPhaseTransitionsFewmonolayer2019, barryModificationSpinicePhysics2019, jaubertSpinIceThin2017, jaubertSpinIcePressure2010, xieMagneticPhaseTransitions2015, edbergDipolarSpinIce2019, piliTopologicalMetamagnetismThermodynamics2022, lu111strainedSpinIce2024}.
We note that further information about the mode structure could be obtained through covariance magnetometry with solid-state defects, a recently developed technique that would be sensitive to the spatial phase variation within a photon mode \cite{rovnyNanoscaleCovarianceMagnetometry2022, cambriaScalableParallelMeasurement2025}.

For ideal superconducting boundaries, the magnitude of the magnetic noise we predict is well within current experimental sensitivities; with intrinsic $T_1$ of over 100~s \cite{jarmolaTemperatureMagneticFieldDependentLongitudinal2012, abobeihOnesecondCoherenceSingle2018, andersenElectronphononInstabilityGraphene2019, cambriaTemperatureDependentSpinLatticeRelaxation2023} and dynamically decoupled $T_2$ times of over 1~s \cite{abobeihOnesecondCoherenceSingle2018, bar-gillSolidstateElectronicSpin2013, zhangPicoteslaFiberizedDiamondbased2024} reported at low temperatures, the additional decoherence rates we predict are easily resolvable (see Fig.~\ref{fig:Fig2} and \ref{fig:Fig3}).
On the other hand, with the ideal insulating boundary conditions, there is no stray-field magnetic noise to measure.
Real samples are not ideal, and may well be governed by some combination of superconducting and insulating boundary conditions -- or have mesoscopic screening lengths at the boundary. 
At low frequencies (where the photon wavelength is long relative to the length scales at the boundary), we expect that such generic boundary conditions would reduce the noise power by a geometric factor of order one relative to the ideal superconducting case, and thus still be observable \cite{SM}. 

Stray-field magnetometry is a strong alternative to more conventional probes in the search for Coulomb phase physics -- if one listens to the magnetic noise, it is indeed possible to ``hear the light''.

We would like to thank A. Chandran, G. Delfino, F. Flicker, D. Long, S. Pace, J. Rau, and S. Zhang for stimulating discussions.
JNH acknowledges support from The Sweden-America Foundation. 
CRL thanks the Max Planck Institute for the Physics of Complex Systems for its hospitality and acknowledges support through the Martin Gutzwiller Fellowship.
This work was in part supported by the Deutsche Forschungsgemeinschaft under grants SFB 1143 (project-id 247310070) and the cluster of excellence ct.qmat (EXC 2147, project-id 390858490).

\bibliography{citations_v3}

\end{document}


\title{Supplementary Material}
\maketitle

\section{Complex scalar on the boundary}
\label{sec:complex_scalar_on_the_boundary}

Pyrochlore quantum spin ice (QSI) in the $U(1)$ spin liquid phase has gapped emergent electric and magnetic charges, both of which are bosonic in nature.
While these charges cost energy in the bulk, they do not cost energy at the boundary for certain terminating cuts of the lattice.
Hence, it is natural to expect that the boundary can be described by a theory of bosons in 2D that are coupled to the gauge fields in the 3D bulk.
Such theories are typically gapped, unless fine-tuned to criticality.
We consider a minimal phenomenological theory of a complex scalar at the boundary described by the action (\cref{eq:Action_Maxwell,eq:Action_boundary} in the main text):
\begin{align}
    {\cal S} &= \int {\rm d}t \int_V {\rm d}^3 x \frac{\hbar}{8\pi \alpha' v} \left(e^2 - v^2 b^2 \right)
    \,,\\
    {\cal S}_{\rm bdy} &= \int {\rm d}t\, {\cal L}_{\rm bdy}=   \int {\rm d}t \int_{\partial V} {\rm d}^2x\, \hbar \left\{ |D_t \psi|^2 - u^2|\vec{D}_{\parallel} \psi|^2- m^2 |\psi|^2 - \frac{\Lambda}{2} |\psi|^4\right\}
    \,.
    \label{eq:sup:action_bdy}
\end{align}
where, $\vec{e}=-\vec{\nabla}\phi - \dot{\vec{a}}$, $\vec{b} = \vec{\nabla}\times\vec{a}$, $D_t=\partial_t+i\,\phi$ and $\vec{D}=\vec{\nabla}-i\,\vec{a}$. For any vector $\vec{A}$, we use the notation $\vec{A}=A_\perp \hat{n}+\vec{A}_\parallel$ to separate the components of the vector perpendicular (along the normal vector $\hat{n}$) and parallel to the boundary.

Variation with respect to $\psi^*$ of the above boundary action gives the equations of motion
\begin{align}
D_t^2\psi-u^2\vec{D}_{\parallel}^2\psi+m^2\psi+\Lambda|\psi|^2\psi=0
\,.
\label{eq:sup:eom_psi}
\end{align}
We are interested in two phases of the boundary theory -- the insulator at $m^2>0$ with $\psi=0$ and the superconductor at $m^2<0$ with $\psi\neq 0$.

The boundary charge and current, $(\rho,\vec{j})$,  can be obtained from the action above,
\begin{align}
    \rho &= -\frac{\delta {\cal L}_{bdy}}{\delta \phi}
    \,,\nonumber\\
    \vec{j} &= \frac{\delta {\cal L}_{bdy}}{\delta \vec{a}} 
    \,.
    \label{eq:sup:bdy_charge_current}
\end{align}
These quantities obey the continuity equation for the boundary charge,
\begin{align}
    \partial_t \rho - \vec{\nabla}\cdot  \vec{j} = 0
    \,.
\end{align}
Variation of the total action with respect to $\phi$ gives:
\begin{align}
    \delta {\cal S}_{\phi} &= \int {\rm d}t \int_V {\rm d}^3 x\,\left(   - \frac{\hbar}{4\pi \alpha' v} e^i\partial_i \delta \phi \right) + \int {\rm d}t \int_{\partial V} {\rm d}^2x\, (-\rho \delta \phi)
    \,.
\end{align}
Using integration by parts in the above expression, we arrive at the equations of motion:
\begin{align}
    \vec{\nabla} \cdot \vec{e} &= 0 &&\text{Bulk} \\
    \hat{n} \cdot \vec{e} &= - \rho/\varepsilon'  && \text{Boundary}
    \label{eq:sup:bdy_phi}
\end{align}
where $\varepsilon' = \frac{\hbar}{4\pi \alpha' v}$, $\hat{n}$ is a unit vector normal to the boundary surface.
Similarly, computing the variation of the total action with respect to $a^i$ gives:
\begin{align}
    \delta {\cal S}_a 
    &= \int {\rm d}t \int_V {\rm d}^3 x\, \delta a^i \frac{\hbar}{4\pi \alpha'} \left(  \frac{1}{v}\,\partial_t e^i    - v\,  [\vec{\nabla}\times \vec{b}]^i  \right) 
    + \int {\rm d}t \int_{\partial V} {\rm d}^2x\, \delta a^i \left(  j^i + \frac{\hbar}{4\pi \alpha'}v [\hat{n} \times \vec{b} ]^i \right)
\end{align}
We obtain,
\begin{align}
    \vec{\nabla} \times \vec{b}- \frac{1}{v^2}\,\partial_t \vec{e} &= \vec{0} &&\text{Bulk}\\
    \hat{n}\times \vec{b}&=-\mu'\,\vec{j} &&\text{Boundary}
    \label{eq:sup:bdy_a}
\end{align}
where $\mu'=4\pi\alpha/(\hbar v)$.

\paragraph{Insulator.} When $m^2>0$, the linearized equation of motion in \cref{eq:sup:eom_psi} has the stationary state $\psi=0$. To describe fluctuations around this stationary state, we can consider the quadratic approximation to the boundary action in \cref{eq:sup:action_bdy} with $\Lambda \simeq0$. 
Integrating out the complex scalar from this action gives:
\begin{align}
 {\cal S}_{\rm bdy}&=   \int {\rm d}t \int_{\partial V} {\rm d}^2 x \frac{\hbar}{2} \left(\chi_e e_\parallel^2 - \chi_b b_\perp^2 \right)
 \,,
\end{align}
where the electric and magnetic susceptibilities, $\chi_e$ and $\chi_b$ can be obtained by a calculation of the one loop diagrams. 
However, from a simple dimensional analysis, we see that 
$\chi_e= C_e\frac{1}{m}$ and $\chi_b= C_b\frac{u^2}{m}$, where, $C_e$ and $C_b$ are dimensionless proportionality constants. Using \cref{eq:sup:bdy_charge_current} and the action above gives boundary charge and current:
\begin{align}
    \rho &= \chi_e \vec{\nabla}\cdot \vec{e}_{\parallel}
    \quad \text{and} \quad
    \vec{j} =  \chi_b\, \vec{\nabla}\times (b_\perp \,\hat{n})+\chi_e\,\partial_t \vec{e}_{\parallel}
    \,.
\end{align}
Using these expressions in \cref{eq:sup:bdy_phi,eq:sup:bdy_a}, we arrive at boundary conditions:
\begin{align}
    e_{\perp} &= - \lambda_e \vec{\nabla}\cdot \vec{e}_{\parallel}
    \,, \nonumber\\
    \hat{n}\times \vec{b}_{\parallel}&=-\lambda_b\, \vec{\nabla}\times (b_\perp \,\hat{n})-\frac{\lambda_e}{v^2} \partial_t \vec{e}_{\parallel}
    \,,
\end{align}
with $\lambda_e=4\pi\alpha'C_e\, \frac{v}{m}$ and $\lambda_b=4\pi\alpha'C_b\, \frac{u^2}{vm}$.
For a photon mode with wave vector $\vec{k}$ and frequency $\omega=vk$, incident on the boundary, these boundary conditions take the form:
\begin{align}
    e_\perp &= - i\lambda_e \vec{k}_\parallel\cdot \vec{e}_{\parallel} 
    \,,\nonumber\\
    \vec{b}_\parallel&=i\lambda_b\,\vec{k}_\parallel b_\perp-i\lambda_e\frac{\omega}{v^2} \,(\hat{n}\times\vec{e}_{\parallel})
    \,.
    \label{eq:sup:nonideal_ins_bdy_condition}
\end{align}
In the limit where the length scales of the microscopic boundary theory are much smaller than the wavelength of the mode we are probing ($\lambda_e,\lambda_b\ll1/k$), we arrive at the boundary conditions in \cref{eq:insulating_bc} of the main text:
\begin{align}
e_\perp=0 \quad \text{and} \quad \vec{b}_\parallel&=\vec{0}
\,.
\label{eq:idealinsulating}
\end{align}
These are identical to the `natural' boundary conditions that would be obtained in the absence of any boundary field $\psi$.

\paragraph{Superconductor.} 
When $m^2<0$, the amplitude of the complex scalar field acquires a mean value
$|\psi|=\sqrt{ -m^2/\Lambda }$.
Separating the phase and amplitude of the complex scalar, $\psi = |\psi| e^{i \vartheta}$, the boundary action in \cref{eq:sup:action_bdy} takes the form
\begin{align}
    {\cal S}_{\rm bdy} =  \int {\rm d}t \int_{\partial V} {\rm d}^2x\, \hbar
    \left\{
    \begin{array}{ll}
    (\partial_t |\psi|)^2 - u^2(\vec{\nabla} |\psi|)^2-m^2 |\psi|^2 - \frac{\Lambda}{2} |\psi|^4  
    +|\psi|^2 \left( (D_t\vartheta)^2 - u^2(\vec{D}_{\parallel}\vartheta)^2 \right)
    \end{array} 
    \right\}
    \,. 
\end{align}
where $D_t \vartheta=(\partial_t\vartheta+\phi)$ and $\vec{D}_\parallel\vartheta=\vec{\nabla}_\parallel \vartheta - \, \vec{a}$.
This action gives
\begin{align}
    \rho  = -2\,\hbar|\psi|^2\, D_t \vartheta \quad \text{and}\quad
    \vec{j}= -2\,\hbar u^2  |\psi|^2\, \vec{D}_{\parallel} \vartheta
    \,.
\end{align}
Taking derivatives of \cref{eq:sup:bdy_phi,eq:sup:bdy_a},  combining the antisymmetric pieces appropriately, and assuming that there are no vortices (i.e. $\partial_i\partial_j\vartheta-\partial_j\partial_i\vartheta=0$ and $\partial_i\partial_t\vartheta-\partial_t\partial_i\vartheta=0$), we arrive at:
\begin{align}
    \frac{1}{4\pi \alpha'}\left( \frac{u^2}{v}\vec{\nabla}(\hat{n} \cdot \vec{e})+v\,\partial_t(\hat{n}\times \vec{b}) \right) &=  -2u^2|\psi|^2\,\vec{e} 
    \,,\nonumber\\ 
    \frac{ v}{4\pi \alpha'}\vec{\nabla} \times(\hat{n}\times \vec{b})&=-2u^2|\psi|^2\,\vec{b} 
    \,.
\end{align}
Rearranging the above equations and simplifying, we arrive at the following boundary conditions:
\begin{align}
     \,\vec{e}_\parallel  &=- \lambda_e\,\vec{\nabla}_{\parallel}\,(e_\perp)-\lambda_b\,\partial_t(\hat{n}\times \vec{b}_\parallel) 
     \,, \nonumber\\ 
    b_\perp &
    =-\lambda_b\, \vec{\nabla}\cdot \vec{b}_\parallel
    \,.
\end{align}
where $\lambda_e=\frac{1}{8\pi \alpha'}\frac{1}{v|\psi|^2}$ and  $\lambda_b=\frac{1}{8\pi \alpha'}\frac{v}{u^2|\psi|^2}$ are the electric and magnetic penetration depths, respectively.
For a photon mode with wave vector $\vec{k}$ and frequency $\omega=vk$, these boundary conditions take the form:
\begin{align}
     \vec{e}_\parallel  &=-i\lambda_e\,\vec{k}_\parallel\,e_\perp
     +i\lambda_b \, \omega(\hat{n}\times \vec{b}_\parallel)  
     \,, \\ 
    b_\perp &=-i \lambda_b\, \vec{k}_\parallel\cdot \vec{b}_\parallel
    \,.
    \label{eq:sup:nonideal_sc_bdy_condition}
\end{align}
In the limit $\lambda_e,\lambda_b\ll1/k$, we recover the boundary conditions in \cref{eq:superconducting_bc} of the main text:
\begin{align}
   \vec{e}_\parallel = \vec{0}
     \quad \text{and} \quad
    b_\perp=0
    \,.
    \label{eq:idealsuperconducting}
\end{align}



\section{Quantization of emergent photon modes in a cavity}    
\label{sec:quantization_of_emergent_photon_modes_in_a_cavity}

In this section, we expand on the modes introduced in \cref{eq:modes} of the main text. For a cuboid cavity, the photon modes can be described by the vector potential,
\begin{align}
\hat{\vec{a}}(\vec{r},t)=\sqrt{ \frac{4}{V} }\sum_{\{\vec{k}\}>0} \sum_{\lambda_{\vec{k}}} \sqrt{\frac{\hbar}{ \varepsilon'\omega_{\vec{k}} }} \,\vec{W}_{\vec{k},\lambda_{\vec{k}}}(\vec{r})\left(\hat{d}_{\vec{k},\lambda_{\vec{k}}} e^{-i\omega t}  + \hat{d}_{\vec{k},\lambda_{\vec{k}}}^\dagger e^{i\omega t}  \right)
\,,
\label{eq:sup:a_modes_in_cuboid}
\end{align}
where $\varepsilon'=\frac{\hbar}{4\pi\alpha'v}$ and $V=L_xL_yL_z$ is the volume of the cavity, and $\hat{d}_{\vec{k},\lambda_{\vec{k}}}$ is the bosonic annihilation operator of the mode specified by $\vec{k}=\pi(n_x/L_x,n_y/L_y,n_z/L_z)$, with $n_x,n_y,n_z\in\mathbb{Z}^+$, and the polarization label $\lambda_{\vec{k}}$.
Note that we choose to work in the temporal gauge ($\phi=0$). The form of $\vec{W}_{\vec{k},\lambda_{\vec{k}}}$ depends on the boundary condition chosen. We shall assume ideal boundary conditions, as given in Eq.~\eqref{eq:idealinsulating} and \eqref{eq:idealsuperconducting}. We comment on what happens to the modes if the boundary conditions are not ideal below. 
\begin{equation}
W^i_{\vec{k},\lambda_{\vec{k}}}(\vec{r})= \left\{
\begin{aligned}
\eta^i_{\vec{k},\lambda_{\vec{k}}} w^{(i)}_{\vec{k}}(\vec{r}) &\quad \quad \text{for ``insulating'' boundaries} \\
\eta^i_{\vec{k},\lambda_{\vec{k}}} u^{(i)}_{\vec{k}}(\vec{r}) &\quad \quad \text{for ``superconducting'' boundaries}
\,,
\end{aligned}\right.
\end{equation}
with
\begin{align}
w^{(x)}_{\vec{k}} (\vec{r}) &=  \sin(k_x r_x)\cos(k_y r_y)\cos(k_z r_z) 
&u^{(x)}_{\vec{k}} (\vec{r})&=  \cos(k_x r_x)\sin(k_y r_y)\sin(k_z r_z) \nonumber \\ 
w^{(y)}_{\vec{k}} (\vec{r})&=  \cos(k_x r_x)\sin(k_y r_y)\cos(k_z r_z)  \qquad\qquad\text{and} 
&u^{(y)}_{\vec{k}} (\vec{r})&=  \sin(k_x r_x)\cos(k_y r_y)\sin(k_z r_z)   \\ 
w^{(z)}_{\vec{k}} (\vec{r})&=  \cos(k_x r_x)\cos(k_y y)\sin(k_z r_z)  
&u^{(z)}_{\vec{k}} (\vec{r})&=  \sin(k_x r_x)\sin(k_y y)\cos(k_z r_z)
\,,\nonumber
\end{align}
where $\vec{\eta}_{\vec{k},\lambda_{\vec{k}}}$ is a unit vector that encodes the polarization of the mode. These functions are orthogonal:
\begin{align}
\int_V {\rm d}^3  r\, w^{(i)}_{\vec{k}}(\vec{r})w^{(i)}_{\vec{k}'}(\vec{r})&=\frac{V}{8} \delta_{\vec{k},\vec{k}'} = \int_V {\rm d}^3  r\, u^{(i)}_{\vec{k}}(\vec{r})u^{(i)}_{\vec{k}'}(\vec{r})
\,,\nonumber \\ 
\int_V {\rm d}^3  r \,\vec{W}_{\vec{k},\lambda_{\vec{k}}} (\vec{r}) \cdot \vec{W}_{\vec{k}',\lambda_{\vec{k}'}'}(\vec{r}) & = \frac{V}{8} \delta_{\vec{k},\vec{k}'} \delta_{\lambda_{\vec{k}},\lambda_{\vec{k}}'}
\,.
\end{align}
One can also check that:
\begin{align}
(\vec{\nabla}\times \vec{W}_{\vec{k},\lambda_{\vec{k}}})^j&= \left\{
\begin{aligned}
-(\vec{k}\times\vec{\eta}_{\vec{k},\lambda_{\vec{k}}})^j u^{(j)}_{\vec{k}}(\vec{r}) &\quad \quad \text{for insulating boundaries}\,,\\
(\vec{k}\times\vec{\eta}_{\vec{k},\lambda_{\vec{k}}})^j w^{(j)}_{\vec{k}}(\vec{r})  &\quad \quad \text{for superconducting boundaries}\,,
\end{aligned}\right.
\,
\end{align}
and
\begin{equation}
    \int_V {\rm d}^3  r \,(\vec{\nabla}\times \vec{W}_{\vec{k},\lambda_{\vec{k}}})\cdot(\vec{\nabla}\times\vec{W}_{\vec{k}',\lambda_{\vec{k}'}'}) = |\vec{k}|^2\frac{V}{8} \delta_{\vec{k},\vec{k}'} \delta_{\lambda_{\vec{k}},\lambda_{\vec{k}}'}
\,.
\end{equation}

The emergent electric and magnetic fields can be obtained from the vector potential:
\begin{align}
\hat{\vec{e}}(\vec{r},t)&=i\sqrt{ \frac{4}{V} }\sum_{\{\vec{k}\}>0} \sum_{\lambda_{\vec{k}}} \sqrt{\frac{\hbar\,\omega_{\vec{k}}}{ \varepsilon' }} \hat{d}_{\vec{k},\lambda_{\vec{k}}} e^{-i\omega t} \,\vec{W}_{\vec{k},\lambda_{\vec{k}}}(\vec{r}) -{\text{h.c.}}  
\,,
\label{eq:sup:e_mode_structure}\\ 
\hat{\vec{b}}(\vec{r},t)&=\sqrt{ \frac{4}{V} }\sum_{\{\vec{k}\}>0} \sum_{\lambda_{\vec{k}}} \sqrt{\frac{\hbar}{ \varepsilon'\omega_{\vec{k}} }} \hat{d}_{\vec{k},\lambda_{\vec{k}}} e^{-i\omega t} \,\vec{\nabla}\times\vec{W}_{\vec{k},\lambda_{\vec{k}}}(\vec{r}) +{\text{h.c.}} 
\,.
\label{eq:sup:b_mode_structure}
\end{align}

Note that the mode quantization is chosen to ensure that the Hamiltonian is diagonal in the cavity boson modes:
\begin{align}
{\cal H} &=\int_V {\rm d}^3 r\, \frac{\varepsilon'}{2}(\hat{e}^2 +v^2 \hat{b}^2)\nonumber \\ 
&= \sum_{\{\vec{k}\}>0} \sum_{\lambda_{\vec{k}}} \hbar\,\omega_{\vec{k}} \left( \hat{d}_{\vec{k},\lambda_{\vec{k}}}^{\dagger}\hat{d}_{\vec{k},\lambda_{\vec{k}}}+\frac{1}{2} \right)
\,.
\label{eq:quantconditions}
\end{align}

\subsection{Counting of modes}\label{sec:counting-of-modes}

There is a discrete set of $\vec{k}$-vectors labeling the modes. 
For each $\vec{k}$, we can have up to two polarization modes, indicated by $\lambda_{\vec{k}}$. The polarization modes can be identified by choosing two unit vectors $\vec{\eta}_{\vec{k},\lambda_{\vec{k}}}$, which are mutually orthogonal, orthogonal to $\vec{k}$, and correspond to two independent cavity modes.
Not all choices of $\vec{k}$ correspond to allowed photon modes in a cuboid cavity. For example, if $\vec{k}$ is chosen to point along the $x,\,y$ or $z$ axis, then $\vec{a}_{\vec{k},\lambda_{\vec{k}}}(\vec{r},t)=\vec{0}$ for any choice of $\vec{\eta}_{\vec{k},\lambda_{\vec{k}}}$. 
If $\vec{k}$ is chosen to lie in the plane formed by two of the three axes (e.g. if $\vec{k} =\pi (1/L_x, 2/L_y,0)$, there is only one choice of $\vec{\eta}_{\vec{k},\lambda_{\vec{k}}}$ for each boundary condition which gives $\vec{a}(\vec{r},t)\neq \vec{0}$). 

A consistent choice of $\vec{\eta}_{\vec{k},\lambda_{\vec{k}}}$ to ensure all modes are accounted for is the following:
\begin{itemize}
\item 
If two components of $\vec{k}$ are zero, there are no modes.
\item 
If only one component, $k_\mu=0$, then there is one polarization mode for each boundary condition:
(i)
``insulating'' has $\vec{\eta}_{\vec{k},\lambda=1}=\hat{\mu}\times \hat{k}$ and
(ii)
``superconducting'' has $\vec{\eta}_{\vec{k},\lambda=2}=\hat{k}\times(\hat{\mu}\times \hat{k})$.
\item 
If all three components of $\vec{k}$ are non-zero, there are two polarization modes for both boundary conditions, and these can be labeled by choosing:
$\vec{\eta}_{\vec{k},\lambda=1}=\hat{z}\times \hat{k}$  and  $\vec{\eta}_{\vec{k},\lambda=2}=\hat{k}\times(\hat{z}\times \hat{k})$.
\end{itemize}



\section{ Zero stray field noise for insulating boundaries}
Let us now formally show that the stray-field magnetic noise is exactly zero for all modes in a cavity with (ideal) insulating boundary conditions.

The emergent electric field is directly related to the magnetization of the sample, $\vec{m}(\vec{r},t)=-g\,\partial_t \vec{a}(\vec{r},t)$, which, when convolved with the dipole kernel, gives the stray magnetic field,
\begin{align}
\hat{B}^{\mu}(\vec{r},t)
  =\int_V {\rm d}^3r'\, H^{\mu \nu}(\vec{r}-\vec{r'})\, \hat{m}^{\nu}(\vec{r'},t)\,
 \qquad \vec{r}\notin V \,.
 \label{eq:sup:B_operator_kernel_integral}
\end{align}
The dipole kernel in the quasi-static limit ($\frac{\omega d}{c}\ll 1$, where $d$ is the distance between the probe and sample) is given by
\begin{align}
    H^{\mu \nu}(\vec{R})
    =- \frac{\mu_0}{4\pi}\,\partial_\mu \partial_\nu \frac{1}{R}
    =\frac{\mu_0}{4\pi}\,
   \frac{\delta^{\mu \nu}R^{2}-3R^\mu R^\nu}{R^{5}}
   \,, \quad \text{for }R>0\,.
   \label{eq:sup:dipolar_kernel}
\end{align}
The stray field noise is given by
\begin{align}
{\cal C}^{\mu \nu}(\vec{r}, \omega) &= \int_{-\infty}^\infty {\rm d} t\, e^{i\omega t} \left\langle \left\{\hat{B}^\mu (\vec{r}, 0), \hat{B}^\nu(\vec{r}, t) \right\} \right \rangle \nonumber \\ 
&=\sum_{\{\vec{k}\}>0} \sum_{\lambda_{\vec{k}}} B_{\vec{k},\lambda_{\vec{k}}}^\mu(\vec{r})B_{\vec{k},\lambda_{\vec{k}}}^\nu(\vec{r}) [2n_{\rm B}(\omega_{\vec{k}}) + 1] 2\pi \delta(\omega-\omega_{\vec{k}})
\,,
\label{eq:sup:noise_definition}
\end{align}
where $n_{\rm B}(\omega)=[\exp{(\beta \hbar \omega)}-1]^{-1}$ is the Bose-Einstein occupation function and
\begin{align}
B_{\vec{k},\lambda_{\vec{k}}}^\mu(\vec{r}) &=\mu_0 g \sqrt{\frac{\hbar \omega_{\vec{k}}}{\varepsilon' V}} \sum_\nu \int_V {\rm d}^3 r' H^{\mu \nu}(\vec{r}-\vec{r}') W^\nu_{\vec{k},\lambda_{\vec{k}}} (\vec{r}') 
\, 
\label{eq:sup:B_kernel_integral}
\end{align}
is the stray magnetic field generated by a specific mode.
The above integral can be equivalently computed in the Fourier space,
\begin{align}
    B^{\mu}_{\vec{k},\lambda_{\vec{k}}}(\vec{r}) &= \mu_0 g \sqrt{\frac{\hbar \omega_{\vec{k}}}{\varepsilon' V}}\int \frac{{\rm d}^3q}{(2\pi )^3}  \widetilde{H}^{\mu\nu}(\vec{q})\, W_{\vec{k},\lambda_{\vec{k}}}^\nu(\vec{q}) \, e^{-i \vec{q}\cdot \vec{r}}  
    \label{eq:FourierSpaceKernel}
\end{align}
with
\begin{align}
\widetilde{H}^{\mu\nu }(\vec{q})&= \int {\rm d}^3 r\  H^{\mu\nu }(\vec{r}) \,e^{i\vec{q}\cdot \vec{r}}= \mu_0 \, \frac{q^\mu q^\nu}{q^2} \label{eq:sup:kernel_in_reciprocal_space}\\ 
\vec{W}_{\vec{k},\lambda_{\vec{k}}}(\vec{q})&=\int{\rm d}^3r\,
e^{i\vec{q}\cdot\vec{r}}\, \vec{W}_{\vec{k},\lambda_{\vec{k}}}(\vec{r})\nonumber \\ 
&=
\left\{
\begin{aligned}
\eta^i_{\vec{k},\lambda_{\vec{k}}} \tilde{w}^{(i)}_{\vec{k}}(\vec{q}) &\quad \quad \text{for insulating boundaries}\\
\eta^i_{\vec{k},\lambda_{\vec{k}}} \tilde{u}^{(i)}_{\vec{k}}(\vec{q}) &\quad \quad \text{for superconducting boundaries}
\end{aligned}\right.
\label{eq:sup:mode_shapes_in_reciprocal_space}
\end{align}
The boundary-dependent Fourier space functions are:
\begin{align}
\tilde{w}^{(x)}_{\vec{k}}(\vec{q}) &= -k_xq_yq_z \, f_{\vec{k}}(\vec{q}) 
& \tilde{u}^{(x)}_{\vec{k}}(\vec{q}) &= -i\,q_xk_yk_z \, f_{\vec{k}}(\vec{q}) 
\nonumber \\ 
\tilde{w}^{(y)}_{\vec{k}}(\vec{q}) &= -q_xk_yq_z \,f_{\vec{k}}(\vec{q})
\qquad \qquad \text{and}  
& \tilde{u}^{(y)}_{\vec{k}}(\vec{q}) &= -i\,k_xq_yk_z \, f_{\vec{k}}(\vec{q}) 
\nonumber \\ 
\tilde{w}^{(z)}_{\vec{k}}(\vec{q}) &= -q_xq_yk_z \, f_{\vec{k}}(\vec{q})
& \tilde{u}^{(z)}_{\vec{k}}(\vec{q}) &= -i\,k_xk_yq_z \, f_{\vec{k}}(\vec{q})\,,
\end{align}
with
\begin{align}
f_{\vec{k}}(\vec{q})=\frac{\left(  1-e^{i(q_x+k_x)L_x} \right)\left(1-e^{i(q_y+k_y)L_y} \right)\left(1-e^{i(q_z+k_z)L_z} \right)}{(k_x^2-q_x^2)(k_y^2-q_y^2)(k_z^2-q_z^2)}.
\end{align}
The noise thus computed for a sample with insulating boundaries is zero, because the integrand in Eq.~\eqref{eq:FourierSpaceKernel} vanishes:
\begin{align}
\widetilde{H}^{\mu\nu}(\vec{q})\, W_{\vec{k},\lambda_{\vec{k}}}^\nu(\vec{q}) &= \mu_0 \frac{q^\mu}{q^2} \sum_\nu q^\nu \eta^\nu_{\vec{k},\lambda_{\vec{k}}} \tilde{w}^{(\nu)}_{\vec{k}}(\vec{q})\nonumber \\ 
&=-\mu_0 \frac{q^\mu}{q^2} q^xq^yq^z f(\vec{q})\, \left(  \vec{\eta}_{\vec{k},\lambda_{\vec{k}}} \cdot \vec{k}\right) = 0 &\quad \quad \text{for insulating boundaries.}
\end{align}
(Remember that the polarization direction $\vec{\eta}_{\vec{k},\lambda_{\vec{k}}}$ is always normal to $\vec{k}$.)



\section{Non-Ideal Boundaries: Microscopic length scales at the boundary}
Let us now return to the phenomenological boundary conditions for the superconducting boundary,
\begin{align}
     \,\vec{e}_\parallel  &=- \lambda_e\,\vec{\nabla}_{\parallel}\,(e_\perp)-\lambda_b\,\partial_t(\hat{n}\times \vec{b}_\parallel) 
     \,, \nonumber\\ 
    b_\perp &
    =-\lambda_b\, \vec{\nabla}\cdot \vec{b}_\parallel
    \, ,
\end{align}
and ask how the modes change if we do not assume that the length scales $\lambda_{e,b}$ are effectively zero. 
For long wavelengths, $\lambda_{e, b}k\ll 1$, the component of the electric field parallel to the boundary turns on as $\vec{e}_\parallel \approx (\lambda_e k + \lambda_b k)\vec{e}_\perp$.
The mode quantization condition, Eq.~\eqref{eq:quantconditions}, tells us that the integrated electric field energy density ($\propto e^2$) is conserved. 
Thus, increasing $\lambda_{e, b}k$ from zero decreases the perpendicular component of the mode approximately according to $e_\perp\sim\sqrt{1-(\lambda_e+\lambda_b)^2k^2}$.
The stray magnetic field is dominated by the magnetization pointing out of the boundary, $M_\perp \propto e_\perp$; relaxing the ideal superconducting boundary conditions (although remaining elastic) reduces the low-frequency stray-field noise by a geometric factor $1-(\lambda_e+\lambda_b)^2\omega^2/v^2$.
In the insulating case the normal component of the magnetization instead turns on as $M_\perp \propto e_\perp \propto (\lambda_e+\lambda_b)\omega/v$, and the stray-field noise thus turns on $\propto (\lambda_e+\lambda_b)^2\omega^2/v^2$.



\section{Ewald Summation to compute stray magnetic field}

In general, the dipole kernel integrals in \cref{eq:sup:B_kernel_integral} are challenging to compute for finite samples; the $1/R^3$ scaling of the kernels and the harmonic variation of the modes lead to slow convergence of the integral. 
In this section, we provide the details of computing this integral using Ewald summation \cite{ewaldBerechnungOptischerUnd1921}. 
This method improves the convergence of the integral by evaluating the short-range and long-range parts of the integral separately in real and reciprocal space, respectively.

Ewald summation makes use of the identity
\begin{equation}\label{eq: erf}
    \frac{1}{R}
     =\underbrace{\frac{\rm{erfc}(\alpha R)}{R}}
      _{\text{short-range part}}
     +\underbrace{\frac{\rm{erf}(\alpha R)}{R}}
      _{\text{long-range part}}
\end{equation}
to separate the dipolar kernel in \cref{eq:sup:dipolar_kernel} into two pieces,
\begin{equation}
    H^{\mu\nu}(\vec{R})=H^{\mu\nu}_{(\mathrm{real})}(\vec{R};\alpha)+H^{\mu\nu}_{(\mathrm{reciprocal})}(\vec{R};\alpha)
\,.
\label{eq:sup:kernel_separation}
\end{equation}
and, subsequently, to separate the integral in \cref{eq:sup:B_kernel_integral} into 
\begin{equation}
    B^{\mu}_s(\vec{r}) = B^{\mu}_s(\vec{r};\alpha)_{(\mathrm{real})} + B^{\mu}_s(\vec{r};\alpha)_{(\mathrm{reciprocal})}
    \,.
    \label{eq:sup:B_sum_real_and_reciprocal}
\end{equation}
Here, $1/\alpha$ is a screening parameter that separates the short-range and the long-range dipolar interactions; these are then separately used to evaluate the dipolar integral in real space and reciprocal space, respectively.
For brevity, we use the label $s=\{\vec{k},\lambda_{\vec{k}}\}$ to specify the mode and the notation $\vec{m}_s (\vec{r})= \vec{W}_s (\vec{r}) \, \mu_0 g \sqrt{{\hbar \omega_{\vec{k}}}/{\varepsilon' V}}$.

The real-space screened kernel is given by 
\begin{align}
    H^{\mu\nu}_{(\mathrm{real})}(\vec{R};\alpha)
      &=\frac{\mu_0}{4\pi}\Bigl[
       \delta^{\mu \nu}\, G_1(R;\alpha)-3\,R^{\mu}R^{\nu}\,G_2(R;\alpha)\Bigr]
       \,,\\[1.2em]
    \text{with}\quad G_1(\vec{R};\alpha) &=
      \frac{\rm{erfc}(\alpha R)}{R^{3}}
     +\frac{2\alpha}{\sqrt\pi}\,
      \frac{e^{-\alpha^{2}R^{2}}}{R^{2}},
    \\[4pt]
    G_2(\vec{R};\alpha) &=
      \frac{\rm{erfc}(\alpha R)}{R^{5}}
     +\frac{2\alpha}{\sqrt\pi}\,
      \frac{e^{-\alpha^{2}R^{2}}}{R^{4}}
     +\frac{4\alpha^{3}}{3\sqrt\pi}\,
      \frac{e^{-\alpha^{2}R^{2}}}{R^{2}}.
\end{align}
Since $\rm{erfc}(x)$ decays extremely fast ($\sim e^{-x^{2}}$), the real space integral, 
\begin{align}
    B^{\mu}_s(\vec{r})_{\mathrm{(real)}}
      &=\int_V{\rm d}^3r'\,
         H^{\mu \nu}_{\mathrm{(real)}}(\vec{r}-\vec{r'}; \alpha)\,
         m^{\nu}_s(\vec{r'})
         \,,
         \label{eq:sup:B_real}
\end{align}
gets most of its contribution from within a region with length scale $1/\alpha$ around the point $\vec{r}$.

The reciprocal-space screened kernel is
\begin{equation}
    H^{\mu \nu}_{(\mathrm{reciprocal})}(\vec{R};\alpha)=-\frac{\mu_0}{4\pi}\,\partial^{\mu}\partial^{\nu}\left(\frac{\mathrm{erf}(\alpha R)}{R}\right)
    \,.
\end{equation}
The Fourier transform of the above equation gives 
\begin{align}
\widetilde H^{\mu \nu}_{(\mathrm{reciprocal})}(\vec{q};\alpha)
  &=
\int {\rm d}^3R\, H^{\mu \nu}_{(\mathrm{reciprocal})}(\vec{R};\alpha) e^{i \vec{q}\cdot\vec{R}} \nonumber\\
&=\mu_0\frac{q^\mu q^\nu}{q^{2}}\,
   e^{-q^{2}/(4\alpha^{2})}
&\text{for } \vec{q}\ne\vec{0}.
\end{align}
The kernel is Gaussian in $\vec{q}$, and will thus pick up contributions only from the long-range part of the integral in real-space.
Note that the Fourier transform is evaluated over the relative coordinate, $\vec{R}=\vec{r}-\vec{r'}$, which runs not only over the sample but also over the region outside the sample where the probe is placed. 
We consider a vacuum box of volume $V_{\mathrm{vac}}=L_{\mathrm{vac}}^3$ that encloses the sample and all the points where the probe is placed. Within this volume, the screened kernel can be approximated as:
\begin{equation}
H^{\mu \nu}_{(\mathrm{recirpocal})}(\vec{R};\alpha)
 \approx \frac{1}{V_{\mathrm{vac}}}\sum_{\vec{G}}
   \tilde{H}^{\mu \nu}_{(\mathrm{reciprocal})}(\vec{G};\alpha)e^{-i\vec{G}\cdot {\vec{R}}}
   \,.
\end{equation}
The spacing between the different values of $\vec{G}$ is set by the size of the vacuum box, $\Delta G = 2\pi/L_{\mathrm{vac}}$.
The reciprocal space integral is now given by:
\begin{align}
B^{\mu}_s(\vec{r};\alpha)_{\mathrm{(reciprocal)}}
  &=\int_V {\rm d}^3r'\,
     H^{\mu \nu}_{\mathrm{(reciprocal)}}(\vec{r}-\vec{r'}; \alpha)\,
     m^{\nu}_s(\vec{r'})
     \,\nonumber\\
&\approx\frac{1}{V_{\mathrm{vac}}}\sum_{\vec{G}}
   \widetilde H^{\mu \nu}_{(\mathrm{reciprocal})}(\vec{G};\alpha)\,
   m^{\nu}_s(\vec{G})\,
   e^{-i\vec{G}\cdot\vec{r}}
   \,,
   \label{eq:sup:B_reciprocal}
\end{align}
where $\vec{m}_s (\vec{G})= \vec{W}_s (\vec{G}) \, \mu_0 g \sqrt{{\hbar \omega_{\vec{k}}}/{\varepsilon' V}}$, with $\vec{W}_s(\vec{G})$ given by \cref{eq:sup:mode_shapes_in_reciprocal_space}.

The exponential suppression from the reciprocal space kernel beyond wavelengths of order $\alpha$  ensures that this integral can be computed for the right choice of $\alpha$.
The size of the vacuum box, $L_{\mathrm{vac}}$, has to be chosen sufficiently greater than the largest linear dimension of the sample to ensure that broadened peaks of $\vec{m}_s(\vec{G})$ are resolved by the $\Delta G$ discretization in the reciprocal space.

\begin{figure}
    \centering
    \includegraphics[width=.6\linewidth]{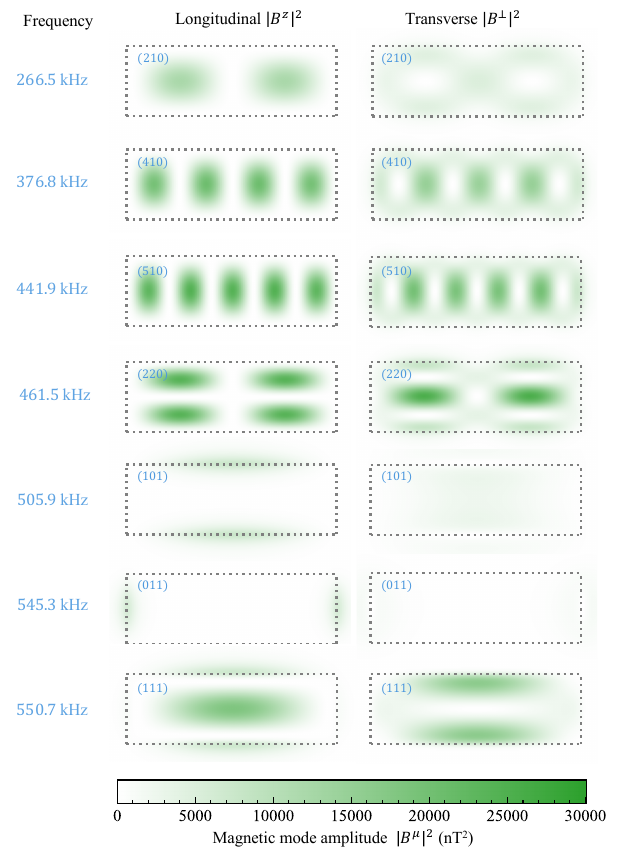}
    \caption{
    Spatially resolved magnetic noise magnitude $|B^z|^2$ and $|B^{\perp}|^2=|B^x|^2+|B^y|^2$ on a plane $0.85\mu$m above a sample with ``superconducting" boundary conditions, of size $65\mu$m$\times22\mu$m and thickness $10\mu$m, for a selection of modes labeled by $(n_x, n_y, n_z)$ and its frequency.
    Converted to experimentally relevant units using $v=10$~m/s, $\alpha'=0.1$, and $\mu_0^2 g^2 = 10^{-38} T^2 m^2 s^2$. These modes are intended to complement those shown in Fig.~\ref{fig:Fig2} of the main text.
     }
    \label{fig:sup:Fig1}
\end{figure}

In \cref{fig:Fig2} of the main text, we presented the spatial variation of the magnetic noise computed using the Ewald summation method for a selection of modes. Here, in \cref{fig:sup:Fig1}, we present the same for a few other modes.



\section{Magnetic Noise over a Thin film}

In this section, we calculate the magnetic noise density from the photon modes within a thin film of QSI by taking the limit $L_x=L_y=L\gg L_z$. 
In the limit $L\to \infty$, there is a continuous set of modes, and we can perform the dipole kernel integrals analytically.

Consider a probe placed over a thin film with its upper surface in the $z=0$ plane. 
The modes within the thin film are labeled by in-plane momentum vectors, $\vec{k}=(k_x,k_y)$, an integer $n$, for the standing waves in the $z$-direction, and a label $\lambda$, for polarization. 
The photon modes are described by the vector potential:
\begin{align}
    &\hat{a}^{\mu}(\vec{r},t) = \int \frac{{\rm d}\omega}{2\pi } e^{-i\omega t}\frac{1}{\sqrt{ L_zL^2 }}\sum_{\vec{k},n\geq 0}e^{i\vec{k}\cdot \vec{\rho}}\sum_{\lambda}  \sqrt{ \frac{\hbar}{ \varepsilon'\omega_{n,\vec{k}}\, } } W^{(\mu)}_{n}(z)\, \hat{a}^{\mu}_{\lambda,n}(\vec{k},\omega)
    \,,\nonumber \\ 
    &\text{with } \hat{a}_{\lambda,n}^{\mu}(\vec{k},\omega)={\left( \eta^{\mu}_{\lambda,n,\vec{k}}\,  \hat{d}_{\lambda,n,\vec{k}}\, \delta(\omega-\omega_{n,\vec{k}}) + s^{(\mu)}\eta^{\mu}_{\lambda,n,-\vec{k}}\,  \hat{d}_{\lambda,n,-\vec{k}}^{\dagger} \,\delta(\omega+\omega_{n,\vec{k}}) \right)}
    \,.
\end{align}
where $s^{(x/y)}=1$, $s^{(z)}=-1$, and $\vec{r}=(\rho_x,\rho_y,z)$.
$\hat{d}_{\lambda,n,\vec{k}}$ and $\hat{d}^\dagger_{\lambda,n,\vec{k}}$ are again the usual boson raising and lowering operators for the modes.
The function $W_n^{(\mu)}(z)$ depends on the boundary condition:
\begin{align}
    W^{(x/y)}_{n}(z) &= \cos\left( \frac{n\pi z}{L_z} \right) \quad\text{and}\quad  W^{(z)}_{n}(z) = i\sin \left( \frac{n\pi z}{L_z} \right) \quad \text{for ``insulating'' boundary,} 
    \label{eq:sup:thin_film_ins_bc}\\    
    W^{(x/y)}_{n}(z) &= \sin\left(  \frac{n\pi z}{L_z}  \right) \quad\text{and}\quad  W^{(z)}_{n}(z) = -i\cos\left(  \frac{n\pi z}{L_z}  \right) \quad \text{for ``superconducting'' boundary} 
    \label{eq:sup:thin_film_sc_bc}
\end{align}
$W^{(z)}$ is chosen to be imaginary to ensure that $\vec{\nabla}\cdot \hat{\vec{a}}(r,t)=0$, or rather, more importantly $\vec{\nabla}\cdot \hat{\vec{e}}(r,t)=0$. The sign factor $s^{(\mu)}$ is required to ensure that $\hat{a}^\mu(r,t)$ is Hermitian.
The magnetization is given by:
\begin{align}
&\hat{m}^{\mu}(\vec{r},t) = -g\partial_t \hat{a}^{\mu}(r,t)\nonumber \\ 
&\,= ig_1\int \frac{{\rm d}\omega}{2\pi } e^{-i\omega t}\frac{1}{\sqrt{ L_zL^2 }}\sum_{\vec{k},n\geq 0}\sum_{\lambda}  \sqrt{ \frac{\hbar\, \omega_{n,\vec{k}}}{ \varepsilon'} }e^{i\vec{k}\cdot \vec{\rho}}\,  W_n^{(\mu)}(z){\left(  \eta^{\mu}_{\lambda,n,\vec{k}}\,  \hat{d}_{\lambda,n,\vec{k}}\, \delta(\omega-\omega_{n,\vec{k}}) - s^{(\mu)}\eta^{\mu}_{\lambda,n,-\vec{k}}\,  \hat{d}_{\lambda,n,-\vec{k}}^{\dagger} \,\delta(\omega+\omega_{n,\vec{k}}) \right)}
\end{align}
The magnetic field outside can be computed with the kernels, whose analytic expressions are known for infinite slabs \cite{chatterjeeDiagnosingPhasesMagnetic2019}:
\begin{align}
&\hat{B}^{\mu}(\vec{r},t)=\mu_0 \int _V {\rm d}^3  r'\, H^{\mu\nu} (\vec{r}-\vec{r}',t) \hat{m}^{\nu}(\vec{r}',t)\nonumber \\ 
&\,=ig_1\mu_0\int \frac{{\rm d}\omega}{2\pi } e^{-i\omega t}\frac{1}{\sqrt{ L_zL^2 }}\sum_{\vec{k},n\geq 0}\sum_{\lambda}  \sqrt{ \frac{\hbar\, \omega_{n,\vec{k}}}{ \varepsilon'} } F^{\mu\nu}_{n,\vec{k}}\, (z,\omega){\left( \eta^{\nu}_{\lambda,n,\vec{k}}\,  \hat{d}_{\lambda,n,\vec{k}}\, \delta(\omega-\omega_{n,\vec{k}}) - s^{(\nu)} \eta^{\nu}_{\lambda,n,-\vec{k}}\,  \hat{d}_{\lambda,n,-\vec{k}}^{\dagger} \,\delta(\omega+\omega_{n,\vec{k}}) \right)}
\end{align}
where,
\begin{align}
F^{\mu\nu}_{n,\vec{k}}\, (z,\omega) &= \int_V {\rm d}^3r'\,H^{\mu\nu}(r-r',\omega) e^{i\vec{k}\cdot \vec{\rho}} W_n^{(\nu)}(z')= \int_{-L_z}^0dz' H^{\mu\nu}_{n,\vec{k}}\, (z-z',\omega) W_n^{(\nu)}(z')
\,.
\end{align}
Here,

\begin{align}
H^{\mu\nu}_{\vec{k}}(z , \omega) &= \frac{e^{-\kappa z}}{2}
\begin{pmatrix}
\frac{\kappa^2 - k_y^2}{\kappa} & \frac{k_x k_y}{\kappa} &  i k_x \\
\frac{k_x k_y}{\kappa} & \frac{\kappa^2 - k_x^2}{\kappa} &  i k_y  \\
i k_x & i k_y & \frac{-k_x^2 - k_y^2}{\kappa}
\end{pmatrix}\,, &\text{with }\kappa=\sqrt{ |\vec{k}|^2-\frac{\omega^2}{c^2} } \,.
\end{align}

With the definition
\begin{align}
Y_n(\kappa)\equiv\int_{-L_z}^0 dz' \, \frac{e^{\kappa  z'}}{2} e^{ i \frac{\pi nz'}{L_z}}  &=\frac{\kappa-i\frac{n\pi}{L_z}}{\kappa^2+\left(\frac{n\pi}{L_z}\right)^2}\left(1 - e^{-i n\pi}e^{-\kappa L_z}  \right) \,,
\end{align}
we arrive at
\begin{align}
F^{\mu(x/y)}_{n,\vec{k}}(z,\omega)&={\rm Re}[Y_n(\kappa)] H^{\mu(x/y)}_{\vec{k}}(z,\omega) \quad\text{and}\quad 
F^{\mu z}_{n,\vec{k}}(z,\omega)={\rm Im}[Y_n(\kappa)] H^{\mu z}_{\vec{k}}(z,\omega)  \quad \text{for ``insulating"\,,} \\
F^{\mu(x/y)}_{n,\vec{k}}(z,\omega)&={\rm Im}[Y_n(\kappa)] H^{\mu(x/y)}_{\vec{k}}(z,\omega)  \quad\text{and}\quad 
F^{\mu z}_{n,\vec{k}}(z,\omega)={\rm Re}[Y_n(\kappa)] H^{\mu z}_{\vec{k}}(z,\omega) \quad \text{for ``superconducting"\,.} 
\end{align}
We now use the anticommutator
\begin{align}
\langle \{\hat{d}_{\lambda,n,\vec{k}},\hat{d}_{\lambda',n',\vec{k}'}^{\dagger}\}\rangle &= \delta_{\lambda,\lambda'}\delta_{n,n'}\delta_{\vec{k},\vec{k}'}\, (2n_B(\omega_{n,\vec{k}})+1)\nonumber \\ 
&=\delta_{\lambda,\lambda'}\delta_{n,n'}\delta_{\vec{k},\vec{k}'}\, \coth\left( \frac{\beta\,\hbar\,\omega_{n,\vec{k}}}{2} \right) \,,
\end{align}
to compute the correlator:
\begin{align}
C^{\alpha,\beta}(r,\omega) &= \int {\rm d}\tau \, e^{i\omega\tau }\langle \{\hat{B}^{\alpha}(r,t+\tau ),\hat{B}^{\beta} (r,t)\}\rangle \nonumber \\ 
&= g^2\mu_0^2 \frac{\hbar }{ \varepsilon' L_zL^2 }  \sum_{\vec{k},n\geq 0}\sum_{\lambda} \omega_{n,\vec{k}} \ F^{\alpha\mu}_{n,\vec{k}}\, (z,\omega)\,F^{\beta\nu}_{n,-\vec{k}}\, (z,\omega)\left(2n_B(\omega_{n,\vec{k}})+1 \right)\nonumber \\ 
&\hspace{8em} {\left( \int \frac{{\rm d}\tau}{(2\pi )^2}\, e^{i(\omega-\omega_{n,\vec{k}})\tau }\,   s^{(\nu)} \eta^{\mu}_{\lambda,n,\vec{k}}\,\eta^{\nu}_{\lambda,n,\vec{k}}  
+ \int \frac{{\rm d}\tau}{(2\pi )^2}\, e^{i(\omega+\omega_{n,\vec{k}})\tau }\,  s^{(\mu)} \eta^{\mu}_{\lambda,n,-\vec{k}}\,\eta^{\nu}_{\lambda,n,-\vec{k}}  \right)}  
\,.
\end{align}
Simplifying the above expression, we are left with
\begin{align}
C^{\alpha,\beta}(r,\omega)&= \frac{g^2\mu_0^2}{2\pi } \frac{\hbar }{ \varepsilon' L_zL^2 }  \sum_{\vec{k},n\geq 0}\sum_{\lambda} \omega_{n,\vec{k}} \ F^{\alpha\mu}_{n,\vec{k}}\, (z,\omega)\,F^{\beta\nu}_{n,-\vec{k}}\, (z,\omega)\left(2n_B(\omega_{n,\vec{k}})+1 \right) \nonumber \\ 
&\hspace{13em}{\left( s^{(\nu)} \eta^{\mu}_{\lambda,n,\vec{k}}\,\eta^{\nu}_{\lambda,n,\vec{k}} \, \delta(\omega-\omega_{n,\vec{k}})
+  s^{(\mu)}  \eta^{\mu}_{\lambda,n,-\vec{k}}\,\eta^{\nu}_{\lambda,n,-\vec{k}} \, \delta(\omega+\omega_{n,\vec{k}}) \right)} 
\,.
\end{align}
We now take the limit $L\to\infty$, to convert the sum over $\vec{k}$ to an integral:
\begin{align}
C^{\alpha,\beta}(r,\omega)&= \frac{g^2\mu_0^2}{2\pi } \frac{\hbar }{ \varepsilon' L_z} \int \frac{{\rm d}^2k}{(2\pi )^2}  \sum_{n\geq 0}\sum_{\lambda} \omega_{n}(\vec{k}) \ F^{\alpha\mu}_{n} (z,\vec{k})\,F^{\beta\nu}_{n} (z,-\vec{k})\Big(2 n_B(\omega_{n}(\vec{k}))+1\Big)\nonumber \\ 
&\hspace{14em} {\left(s^{(\nu)} \eta^{\mu}_{\lambda,n}(\vec{k})\,\eta^{\nu}_{\lambda,n}(\vec{k}) \, \delta(\omega-\omega_{n}(\vec{k}))
+  s^{(\mu)}  \eta^{\mu}_{\lambda,n}(-\vec{k})\,\eta^{\nu}_{\lambda,n}(-\vec{k}) \, \delta(\omega+\omega_{n}(\vec{k})) \right)} 
\,.
\end{align}
This integral can be performed by using the delta functions. For $\omega>0$, only the first delta function gives a contribution,
\begin{align}
C^{\alpha,\beta}(r,\omega)
&= \frac{g^2\mu_0^2}{(2\pi)^3 } \frac{4\pi \alpha'v }{ L_z}\left(2n_B(\omega)+1 \right)\sum_{n= 0}^{N(\omega)}\sum_{\lambda} \int_{S_n} {\rm d} k\,   \frac{\omega^2}{v^2|\vec{k}|} \ F^{\alpha\mu}_{n,\vec{k}}\, (z,\omega)\,F^{\beta\nu}_{n,-\vec{k}}\, (z,\omega)\,s^{(\nu)} \eta^{\mu}_{\lambda,n,\vec{k}}\,\eta^{\nu}_{\lambda,n,\vec{k}}
\,,
\end{align}
where, $S_n$ is a circle of radius $|\vec{k}|=\sqrt{\frac{\omega^2}{v^2}- \left(\frac{n\pi}{L_z} \right)^2}$ and $N(\omega) = \left\lfloor \frac{|\omega|}{v} \right\rfloor$.



\section{XY8-N pulse sequences and decoherence times}
In \cref{fig:Fig2} of the main text, we illustrate the discrete stray-field magnetic noise spectrum generated by a finite sample of QSI by computing the magnetic noise induced XY8-4 decoherence time, $T_2$, of a nitrogen vacancy center outside the sample. 
This type of ``$T_2$-spectroscopy'' is a standard technique for probing magnetic noise spectra using solid state defects \cite{degenQuantumSensing2017,rovnyNanoscaleDiamondQuantum2024}. 
``XY8-4'' indicates that the measurement is performed during a dynamical decoupling sequence -- a set of microwave $\pi$-pulses rotating the spin about the $x$ and $y$ axes. The pulses are equally spaced in time, with pulse spacing $\tau_{\rm XY8}$, and the basic XY8 building block consists of an eight pulse sequence $XYXYYXYX$.
The ``-4'' denotes that this sequence is repeated four times, for a total of 32 microwave pulses. 
Dynamical decoupling makes the probe sensitive primarily to frequencies around $\pi/\tau_{\rm XY8}$, filtering out noise at other frequencies.

More precisely, the decoherence time is given by 
\begin{equation}
    T_2^{-p} = \frac{\gamma_e^2}{\pi (N \tau_{\rm XY8})^p} 
    \int_0^\infty {\rm d} \omega\  {\cal C}^{zz} \frac{F_N(\omega \tau_{\rm XY8})}{\omega^2} 
    \, ,
\end{equation}
where $p>0$ is the stretching factor (typically close to 1) and $\gamma_e= -2\pi \times 28 $~GHz T$^{-1}$ is the gyromagnetic ratio.
$N$ is the total number of pulses (32 in our case), and the filter function takes the form \cite{rovnyNanoscaleDiamondQuantum2024}:
\begin{equation}
    F_N(\omega \tau) = 8 \sin^4{\left(\frac{\omega \tau}{4}\right)} \frac{\sin^2{\left(\frac{N \omega \tau }{2}\right)}}{\cos^2{\left(\frac{\omega \tau}{2}\right)}}
    \, .
\end{equation}
The filter function has a primary peak at $\omega=\pi/\tau_{\rm XY8}$, and this peak narrows with increasing pulse number $N$. 
There are additional, smaller peaks at $\omega=3\pi/\tau_{\rm XY8},\ 5\pi/\tau_{\rm XY8},\ ...$, which loose weight as $N$ increases. 
These additional peaks are the cause of the fluctuations seen in Fig.~\ref{fig:Fig2}a of the main text at frequencies below the first cavity mode and between the mode peaks.



\bibliography{citations_v3}

\newpage
